\documentclass[twocolumn,letterpaper,aps,prc,superscriptaddress,nofootinbib,showpacs,floatfix]{revtex4-1}
\usepackage{graphicx}
\usepackage{comment}
\usepackage{setspace}
\usepackage{amsmath}
\usepackage{amssymb}
\usepackage{color}
\usepackage{indentfirst}
\usepackage{xspace}
\usepackage{hyperref}
\usepackage{verbatim}
\usepackage{epstopdf}

\usepackage{lineno}

\begin{document}


\title{Comprehending particle production at RHIC and LHC energies using global measurements}
 \author{Sadhana Dash} \author{Basanta K. Nandi} \author{Ranjit Nayak} \author{Ashutosh Kumar Pandey}
\author{Priyanka Sett}
\
\affiliation{Indian Institute of Technology Bombay, Mumbai, India}

\email{sadhana@phy.iitb.ac.in}

\date{\today}

\begin{abstract}
The centrality dependence of the charged-particle multiplicity densities 
($dN_{ch}/d\eta$) and  transverse energy densities ($dE_{T}/d\eta$)
are investigated  using the two-component Glauber approach 
for  broad range of energies in heavy ion  collisions at RHIC and
LHC. A comprehensive study shows that the data is well described 
within the framework of two component model which includes the
contribution of ``soft processes'' and ``hard processes''  for
different centrality classes and energies. The data at two different
energies are compared by means of the ratio of $dN_{ch}/d\eta$ (and
$dE_{T}/d\eta$)  to  see the interplay of energy and $x$-scaling.  

\end{abstract}

\maketitle

\section{Introduction}
Heavy ion collisions at relativistic energies provide means to study
strongly-interacting matter are very high densities and temperatures.
The charged particle multiplicity distribution has been
one of the most widely studied experimental observable to comprehend the
hot and dense matter created in such collisions and in general to
understand the particle production mechanism  in such systems. 
Another important observable which gives similar insight is the
distribution of the transverse energy, $E_{T}$ in such
collisions. Experimentally,  the aforementioned global distributions
are quantified  by the charged particle multiplicity density
($dN_{ch}/d\eta $) and the transverse energy density ($dE_{T}/d\eta$). 
 In heavy ion collisions, ``hard" scattering processes, characterized by the
production of high transverse momentum ($p_{T}$)  particles, are
expected to scale with the number of collisions, $N_{coll}$ while, the
production of low $p_{T}$  particles (from soft processes)  is
expected to scale with the number of participants, $N_{part}$ \cite {twocomp}. 
The dependence of the $dN_{ch}/d\eta $  and $dE_{T}/d\eta$  with collision
energy and centrality  gives important information on the relative
contribution of hard and soft processes to the particle production
mechanism.
In this paper, we have used three approaches to study the variation of
$dN_{ch}/d\eta $  (and $dE_{T}/d\eta$) with centrality and beam
energy. The first approach is inspired by the two-component models  
which decompose nucleus-nucleus collisions  into  soft  and  hard
interactions,  where  the  soft  interactions  produce  particles
with  an average multiplicity proportional to  $N_{part}$  
 and the probability for hard interactions to occur is proportional to
 $N_{coll}$ \cite{twocomp}.
The two component approach \cite{twocomp} parametrizes
$dN_{ch}/d\eta $ by the following expression.
\begin{equation}
  \frac{dN_{ch}}{d\eta} = n_{pp} \, \left( (1-x) \, \frac{\langle N_{part} \rangle}{2} + x \, \langle N_{coll} \rangle \right)
  \label{e1}  
\end{equation}  
 In this model,  $n_{pp}$ represents the average charged-particle multiplicity in a
single nucleon-nucleon collision,  while $x$ refers to the 
fraction of ``hard" partonic interactions.
In a previous study~\cite{mitpaper}, the variation of $dN_{ch}/d\eta $ was described
by using the above model where both  $n_{pp}$  and $x$ were treated as free parameters. 
As $n_{pp}$ represents a physical observable (mean multiplicity in
$pp$ collisions),  the above approach might not give the proper
values and dependence of  $x$ and $n_{pp}$  with collision energy and
centrality. The obtained values for the fit  parameters might be an
artifact of the fitting procedure.
The method can be improvised by constraining the parameter $ n_{pp}$ by
using the experimentally measured values at a given energy.  This
would result in a robust extraction of $x$ parameter with respect to
centrality and beam energy.
In other two approaches,  $dN_{ch}/d\eta $  was studied by single
component parameterization \cite{sps1, sps2, sps3, alice1}.  
A similar approach was also used to explain the variation of
$dE_{T}/d\eta$ dependence with collision centrality and energy.
In the case of $dE_{T}/d\eta$ , we do not have a broad experimental
measurement of $E_{T_{pp}}$  (for nucleon nucleon collisions) and hence 
mean $E_{T_{pp}}$  was treated as a free parameter. We have  used the
 $x$ parameter values obtained from the  $dN_{ch}/d\eta $ fit. This
assumption  seems valid as the $x$ value would remain same 
for similar centralities and energies.  
This method can provide an estimation of variation of  $E_{T_{pp}}$
with collision energy.

\begin{figure}[htb]
\begin{center}
\includegraphics[scale=0.4]{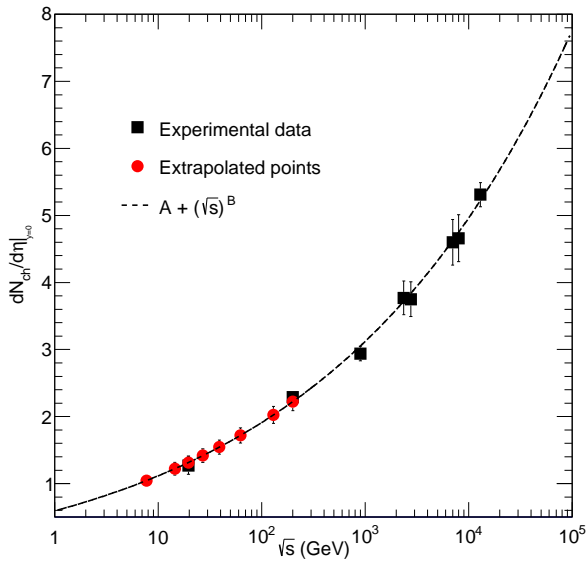}
\caption{The charged particle multiplicity density 
measured in $pp$ collisions at  RHIC \cite{npp_rhic} and  LHC \cite{npp_lhc1, npp_lhc2} energies. The solid squares are the experimentally measured $n_{pp}$ data.  
The data is described by a power law of the form $A + \sqrt{s}^{\,B}$, shown by the dashed line. The solid circles are the extrapolated 
data for the RHIC beam scan energies.}
\label{f1}
\end{center}
\end{figure}

\section{Analysis}

The data for  $dN_{ch}/d\eta $  and $dE_{T}/d\eta$  for Au+Au
collisions at mid-rapidity have been presented by  PHENIX experiment
for a large set of energies at RHIC \cite{phenixpaper}. The LHC data for $dN_{ch}/d\eta $  in Pb+Pb collisions
at 2.76 TeV and 5.02 TeV has been taken from ALICE experiment \cite{alice1,alice2} and the $dE_{T}/d\eta$ data for 2.76 TeV 
can be found in Ref.~\cite{alice3} .
As discussed in previous section, the values of $n_{pp}$ were
taken to be the experimentally measured values for a particular
energy. Since we do not have experimental measurement of  charged-particle 
multiplicity in $pp$ collisions for most of the beam energy scan 
energies at RHIC, the values for $n_{pp}$,  were obtained by extrapolating the 
 $\sqrt{s}$  dependence of measured  $n_{pp}$  in $pp$ collisions at different energies. 
 This is  depicted in Figure~\ref{f1}. The measured experimental 
 data from RHIC \cite{npp_rhic} and LHC \cite{npp_lhc1, npp_lhc2} are shown by the solid squares. 
 The data is described by a power law shown by the dashed line. The 
 closed circles are the estimated $n_{pp}$ values for the beam scan 
 energies at RHIC.  
To compare the two-component model with the measured data, one
also needs the values of $\langle N_{part} \rangle$  
and $\langle N_{coll} \rangle$.
The values of $\langle N_{part} \rangle$ are obtained from the
Monte-Carlo Glauber model \cite{glauber} which simulates the heavy ion
collision process on an event by event basis.
The values of $\langle N_{coll} \rangle$  can be obtained by a simple
power law parameterization of $\langle N_{part} \rangle$ obtained from the Monte-Carlo Glauber model \cite{glauber}.
The dependence is given by, 
$N_{coll}$  =  $ a \times {N_{part}}^{b}$. 
The dependence of the parameters for various collision energies is
shown in Table~\ref{ncollnpart}. The value of the parameter $x$ is obtained from the fit to the data employing Eq.~\ref{e1}.  
\begin{table}[!htb]
\centering
\begin{tabular}{|c|c|c|}
\hline
\hline
 $\sqrt{s_{NN}}$ & $a$  & $b$ \\
\hline
7.7 GeV & 0.315 $\pm$ 0.099 & 1.34 $\pm$ 0.057 \\
19.6  GeV &  0.312 $\pm$ 0.096 &  1.35 $\pm$ 0.055 \\
27 GeV & 0.306 $\pm$ 0.1 & 1.36 $\pm$ 0.058 \\
39 GeV & 0.301 $\pm$ 0.096 & 1.36 $\pm$ 0.056 \\
62.4 GeV & 0.395 $\pm$ 0.097 & 1.32 $\pm$ 0.054 \\
130 GeV  & 0.367 $\pm$ 0.144 & 1.35 $\pm$ 0.073 \\
200 GeV & 0.356 $\pm$ 0.117 & 1.36 $\pm$ 0.065 \\
2.76 TeV & 0.365 $\pm$ 0.052 & 1.41 $\pm$ 0.025          \\
5.02 TeV & 0.352 $\pm$ 0.523 &  1.42 $\pm$ 0.027 \\
\hline
\hline
\end{tabular}
\caption{ \label{ncollnpart} The parameters $a$ and $b$ obtained from the 
parameterization of $N_{part}$ and $N_{coll}$ for different collision energies. } 
\end{table}

\begin{figure}
\includegraphics[scale=0.4]{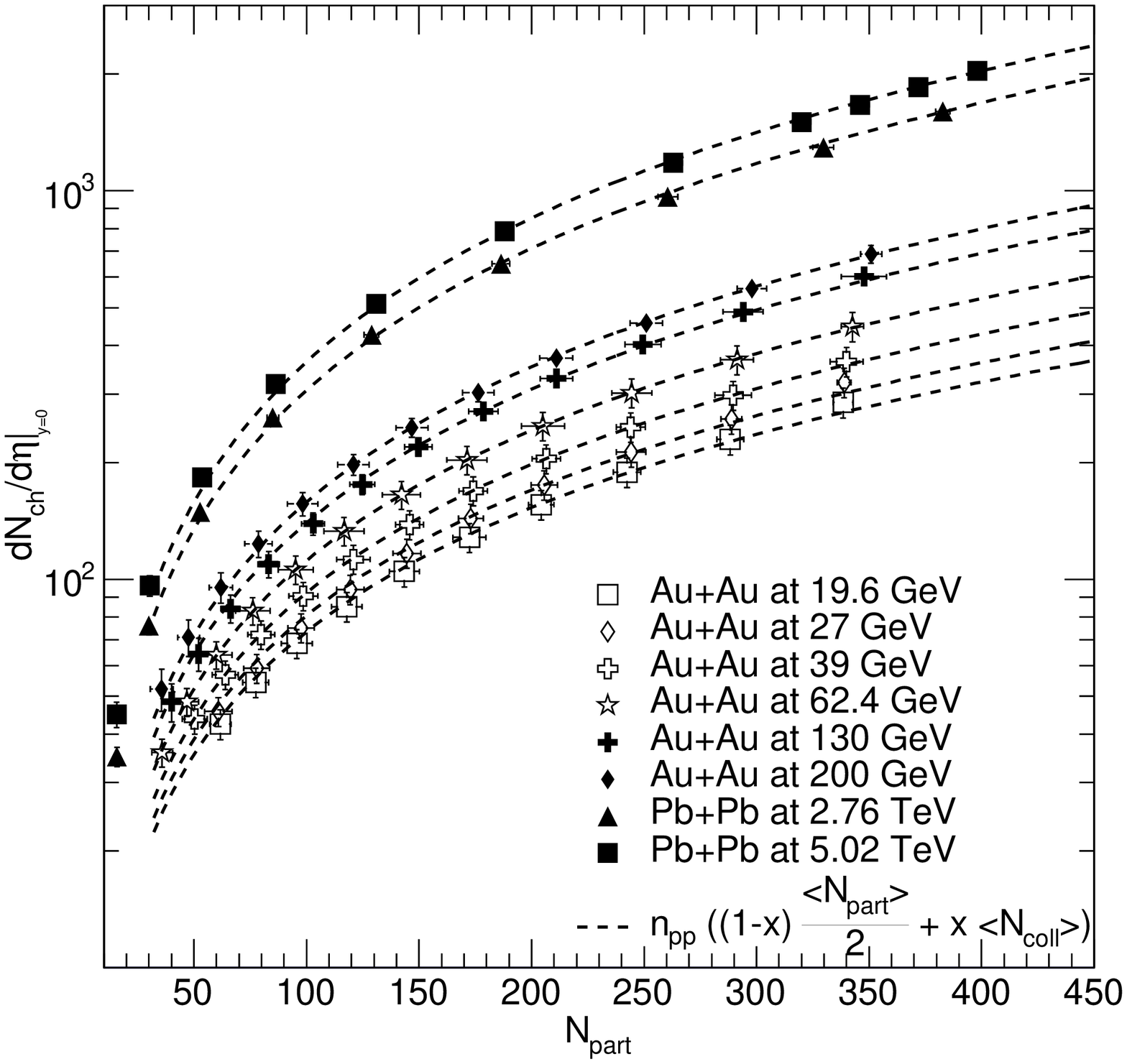}
\caption{Charged particle multiplicity densities measured in 
heavy ion collisions at RHIC (Au+Au)~\cite{phenixpaper} 
and LHC (Pb+Pb)~\cite{alice1, alice2} energies. The dashed line 
represents the fits to Eq.~\ref{e1}.}
\label{f2}
\end{figure}

\begin{figure}
\includegraphics[scale=0.4]{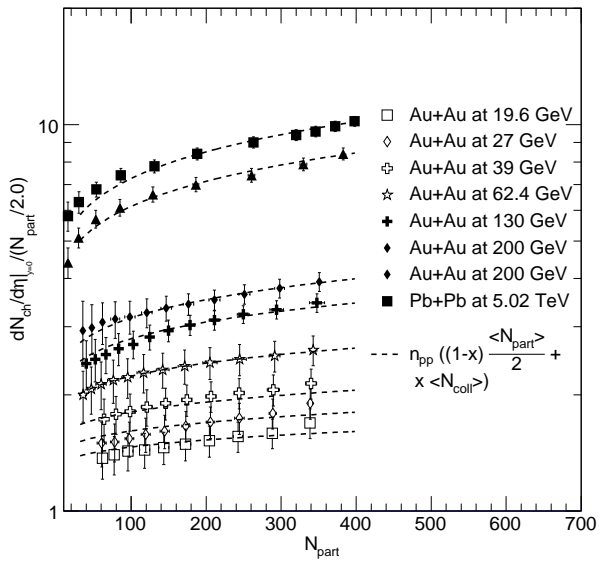}
\caption{Charged particle multiplicity densities scaled by the 
number of participants measured in 
heavy ion collisions at RHIC (Au+Au)~\cite{phenixpaper} 
and LHC (Pb+Pb)~\cite{alice1, alice2} energies. The dashed line 
represents the fits to Eq.~\ref{e1}.}
\label{f3}
\end{figure}

The $dN_{ch}/d\eta $ has also been described by 
single parameter of the following forms~\cite{sps1, sps2, sps3, alice1}, 
\begin{equation}
dN_{ch}/d\eta \propto  \langle N_{part} \rangle ^{\alpha} 
\label{e2}
\end{equation}
\begin{equation}
dN_{ch}/d\eta \propto  \langle N_{coll} \rangle ^{\beta}  
\label{e3}
\end{equation}
The analysis of  $dE_{T}/d\eta$  was also carried in a similar
manner with a slight modification. In this case, the values of  $x$ parameter
obtained from the $dN_{ch}/d\eta $ (Eq.~\ref{e1}) fits were used
while the the mean transverse energy obtained in $pp$ collision was treated as a free parameter. 
This is done as we do not have published experimental results
of mean transverse energy in $pp$ collisions at different energies. 
The above approach would also give us the energy dependence of the
mean transverse energy in $pp$ collisions ($E_{T_{pp}}$).
In addition, the pairwise ratios between the $dN_{ch}/d\eta $ 
and $dE_{T}/d\eta$  per participant pairs for two different  energy
sets were also evaluated. The ratios were calculated 
by interpolating the values of charged particle densities at a particular energy to the
same  $N_{part}$ values at a different energy.
\begin{figure}
\includegraphics[scale=0.4]{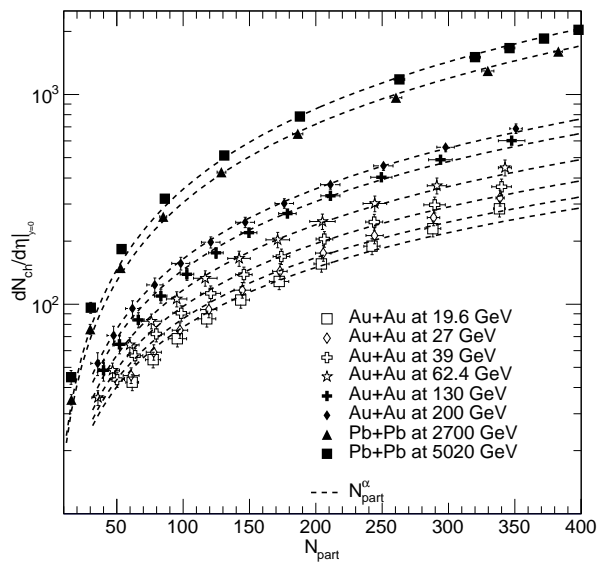}
\caption{Charged particle multiplicity densities measured in 
heavy ion collisions at RHIC (Au+Au)~\cite{phenixpaper} 
and LHC (Pb+Pb)~\cite{alice1, alice2} energies. The dashed line 
represents the fits to Eq.~\ref{e2}.}
\label{f4}
\end{figure}

\begin{figure}
\includegraphics[scale=0.4]{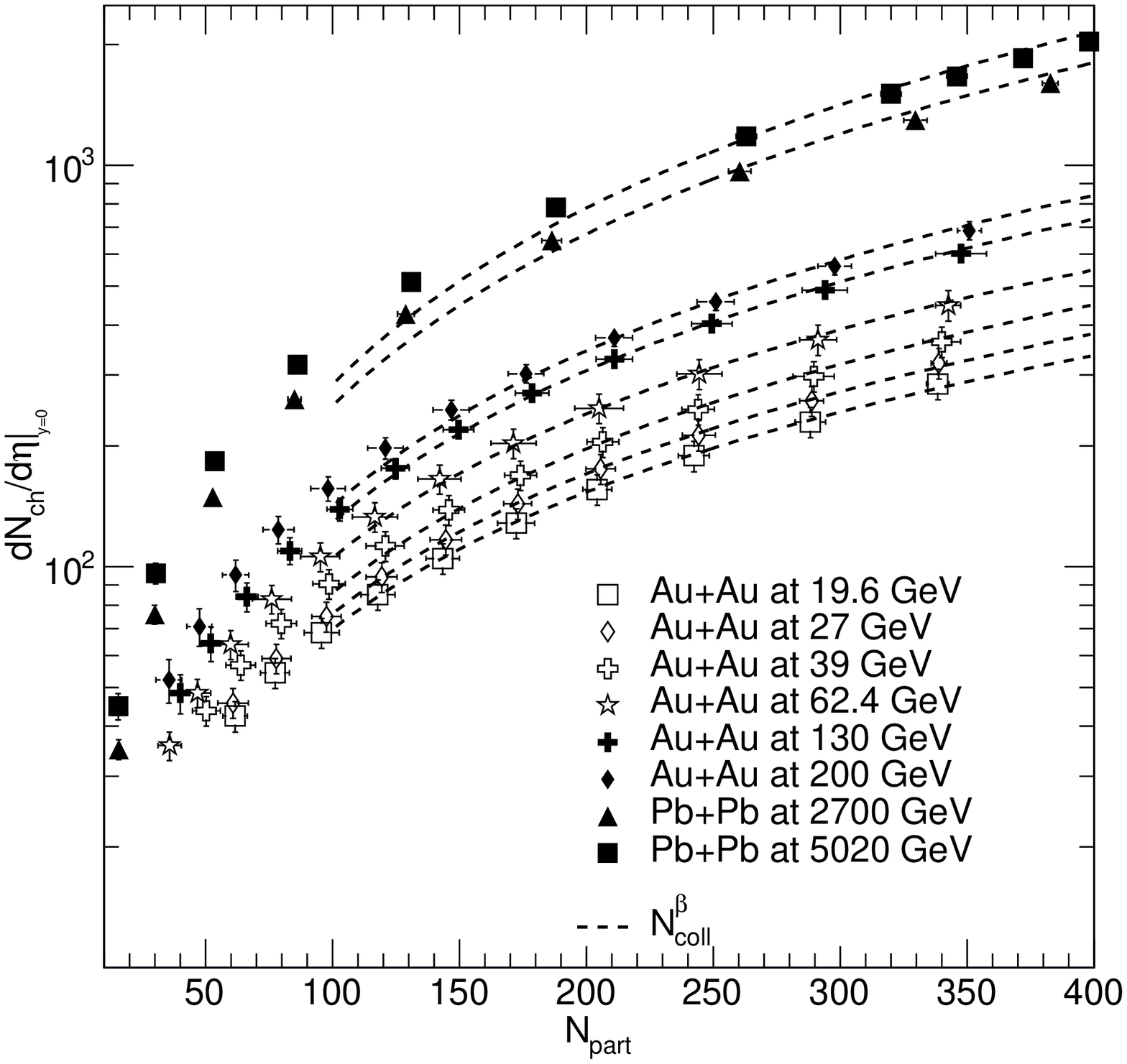}
\caption{Charged particle multiplicity densities measured in 
heavy ion collisions at RHIC (Au+Au)~\cite{phenixpaper} 
and LHC (Pb+Pb)~\cite{alice1, alice2} energies. The dashed line 
represents the fits to Eq.~\ref{e3}.}
\label{f5}
\end{figure} 

\begin{figure}
\includegraphics[scale=0.4]{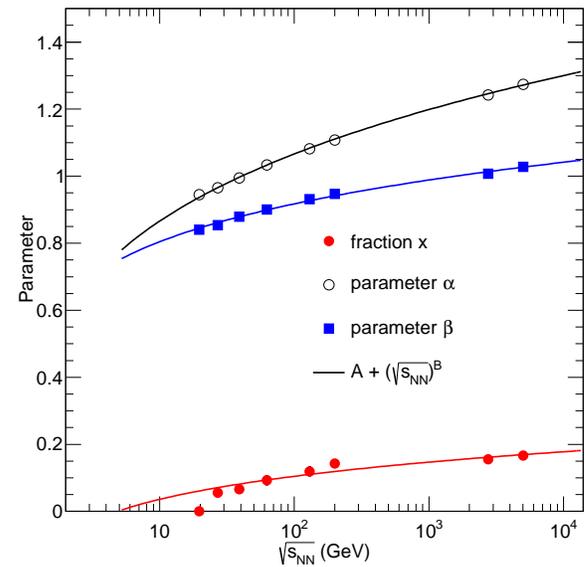}
\caption{The variation of the parameters $x$ (solid circles),
 $\alpha$ (open circles) and $\beta$ (solid squares)  
as a function of $\sqrt{s_{NN}}$. The solid lines are the power 
law parameterization. }
\label{f6}
\end{figure}

\begin{figure}
\includegraphics[scale=0.4]{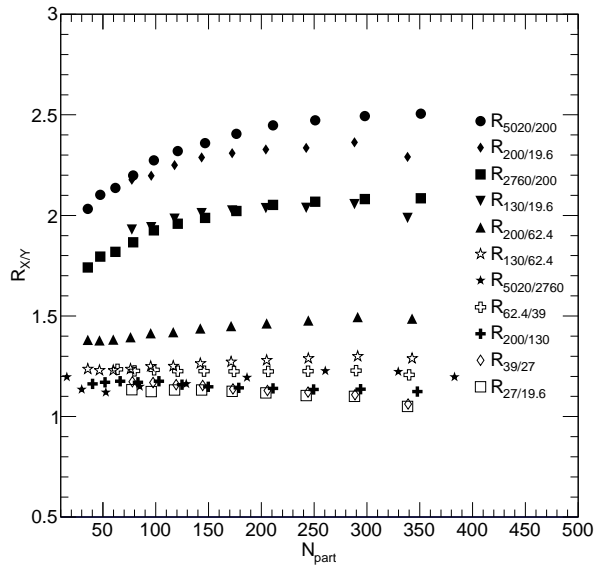}
\caption{The variation of the double ratio of the 
charged particle multiplicity density (refer text) as a function of $\langle N_{part} \rangle$. }
\label{f7}
\end{figure}

\section{Results : Charged particle multiplicity densities}
The  results obtained  from the two component (Eq.~\ref{e1}) fits of
$dN_{ch}/d\eta $  measured at  RHIC (\cite{phenixpaper}) 
and LHC (\cite{alice1, alice2}) energies are shown in
Figure~\ref{f2}.  The dashed lines show the two-component fits performed by $\chi^{2}$ minimization.  

Figure~\ref{f3} shows the same data but the multiplicity densities are
scaled with the number of participant pairs.
One can observe from Figure~\ref{f3}, that at RHIC energies, the 
scaled multiplicity densities do not have a strong dependence on the 
$\langle N_{part} \rangle$ and have a flat behavior. However, the observed trend
for LHC differs from  RHIC .  As a function of system size, 
the scaled multiplicity density shows  an increasing trend, which one can 
expect from the large difference in collision energies.

The fits for the single component parameterizations,  Eq.~\ref{e2} and 
Eq.~\ref{e3} are shown in Figure~\ref{f4}  and Figure~\ref{f5}, respectively. 
Figure~\ref{f4} shows a good agreement of the parameterization 
with the measured data for all energies.  While using the parameterization stated in Eq.~\ref{e3}, 
we restrict the fit in the common $\langle N_{part} \rangle$ region to have a consistent approach.  

All three parameters, namely $x$, $\alpha$ and $\beta$ are also 
studied as a function of $\sqrt{s_{NN}}$ and are shown in Figure~\ref{f6}. 
One can observe an increasing trend with energy.  
This is  expected as  the number of hard scattering processes 
increase with an increase in number of collisions which increases with energy.
The increasing trend can be analytically described by a power law, 
$A + \sqrt{s_{NN}}^{\,B}$. 
Table \ref{xalphabeta} summarizes the value of the parameters extracted from the fit.

\begin{table}
\centering
\begin{tabular}{|c|c|c|c|}
\hline
\hline
 $\sqrt{s_{NN}}$ & $x$           & $\alpha$           & $\beta$ \\
\hline
19.6  GeV &  0.056 $\pm$ 0.013   & 0.95 $\pm$ 0.006   & 0.84$\pm$ 0.006 \\
27 GeV & 0.066 $\pm$ 0.013       & 0.97 $\pm$ 0.006   & 0.85$\pm$ 0.006\\
39 GeV & 0.093 $\pm$ 0.014       & 0.99 $\pm$ 0.006   & 0.88$\pm$ 0.006\\
62.4 GeV & 0.12 $\pm$ 0.014      & 1.03 $\pm$ 0.006   & 0.90$\pm$ 0.006\\
130 GeV  & 0.14 $\pm$ 0.009      & 1.08 $\pm$ 0.006   & 0.93$\pm$ 0.004\\
200 GeV & 0.16 $\pm$ 0.009       & 1.11 $\pm$ 0.004   & 0.94$\pm$ 0.004\\
2.76 TeV & 0.17 $\pm$ 0.006      & 1.24 $\pm$ 0.003   & 1.01$\pm$ 0.003\\
5.02 TeV & 0.18 $\pm$ 0.004      & 1.27 $\pm$ 0.002   & 1.02$\pm$ 0.002\\
\hline
\hline
\end{tabular}
\caption{ \label{xalphabeta} The parameters $x$,  $\alpha$ and $\beta$ obtained from the fits of multiplicity density distributions using Eq.~\ref{e1}, Eq.~\ref{e2} and Eq.~\ref{e3} for different collision energies. } 
\end{table}

We also compare the pair wise ratios at two different set of 
energies as shown in Figure~\ref{f7}. 
A variable $R_{X/Y} \, = \frac{dN/d\eta}{\langle N_{part}\rangle/2}|_X/\frac{dN/d\eta}{\langle N_{part}\rangle/2}|_Y
$,  is used to denote the magnitude of the 
ratio of multiplicity densities at energy $X$ and energy $Y$ 
(where $X > Y$ ). 
The ratio is evaluated for similar $\langle N_{part} \rangle$  values
for both the energies considered~\cite{mitpaper}. 
The ratio shows a flat behavior as a function of system size for 
collision energy ranges which differs by a factor of  3 or less i.e. $R_{27/19.6}$, $R_{39/27}$,  
$R_{62.4/39.0}$ and  $R_{200/130}$. 
However, as the value of the collision energy  differs by a factor of
 4 or more, the magnitude of the ratio increases, as well as, the
 ratio shows an increasing trend as a function of $\langle N_{part}
\rangle$.  There is an increase of $\sim$ 20 \%  from most peripheral
to most central collisions.  
On careful observation, one can see that the trend does not
follow strictly the energy difference only. For example,  the values of
$R_{2760/200}$ is consistently lower than $R_{200/19.6}$ although the
collision energy  difference is more for the former than the
later. Therefore, one should also take into account an increase in the $x$ factor
along with an increase of energy. The interplay of both these factors i.e
the energy difference and the increased contribution of  the hard scattering component  
determines the observed trend. The  contribution from both the factors
are tabulated in   \ref{tratio}.
\begin{table}
\centering
\begin{tabular}{|c|c|c|c|c|}
\hline
\hline
 $X$ (GeV) &  $Y$ (GeV)  & $R_{\sqrt{s_{NN}}}$     &   $R_x $     & $R_{\sqrt{s_{NN}}} \times R_x$ \\
  & & $= X/Y$  & $=  X_{x}/Y_{x}$ & \\
\hline
5020 & 200 & 25.1 & 1.13 & 28.39 \\
200 & 19.6  & 10.2 & 2.77 & 28.27 \\
130 & 19 & 6.63 &  2.55 & 16.9 \\
2760 & 200 & 13.8 & 1.07 & 14.8 \\
200 & 62 & 3.21 & 1.3 & 4.18 \\
130 & 62 &  2.08 & 1.2 & 2.49 \\
62 & 39  & 1.6 & 1.29 & 2.06 \\
39 & 27 & 1.44 &  1.41 & 2.03 \\
5020 & 2760 & 1.82 & 1.06 & 1.92 \\
200 & 130 & 1.54 & 1.09 & 1.67 \\
27 & 19 &1.378 & 1.17 & 1.61 \\
\hline
\hline
\end{tabular}
\caption{ \label{tratio} 
The ratio of $X/Y$ and the ratio of the $x$ fraction of the 
corresponding center of mass energies.} 
\end{table}

\section{Results : Transverse energy density}

\begin{figure}
\includegraphics[scale=0.4]{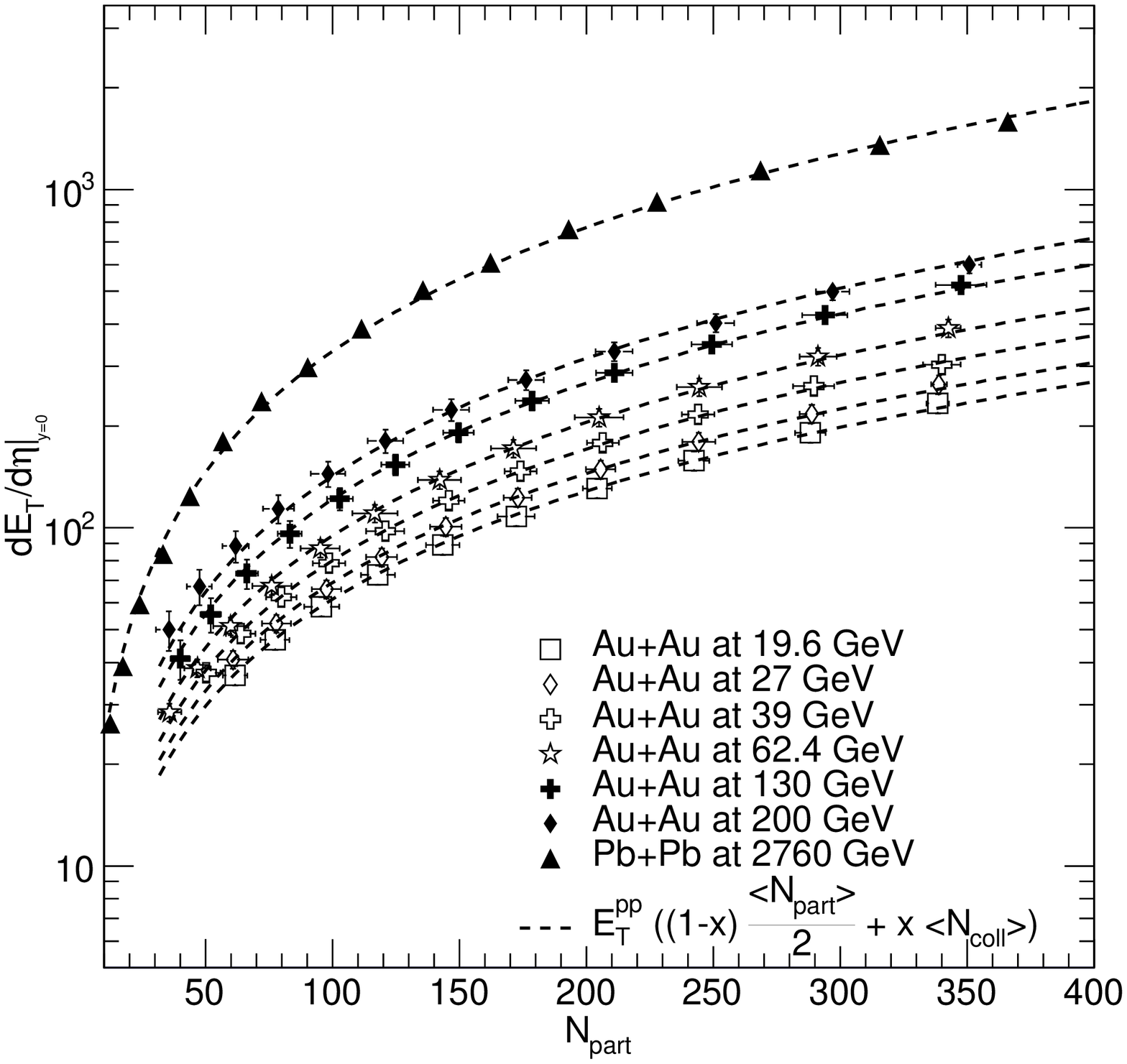}
\caption{Charged particle transverse energy densities measured in 
heavy ion collisions at RHIC (Au+Au)~\cite{phenixpaper} 
and LHC (Pb+Pb)~\cite{alice3} energies. The dashed line 
represents the fits to Eq.~\ref{e1}.}
\label{f8}
\end{figure}

\begin{figure}
\includegraphics[scale=0.4]{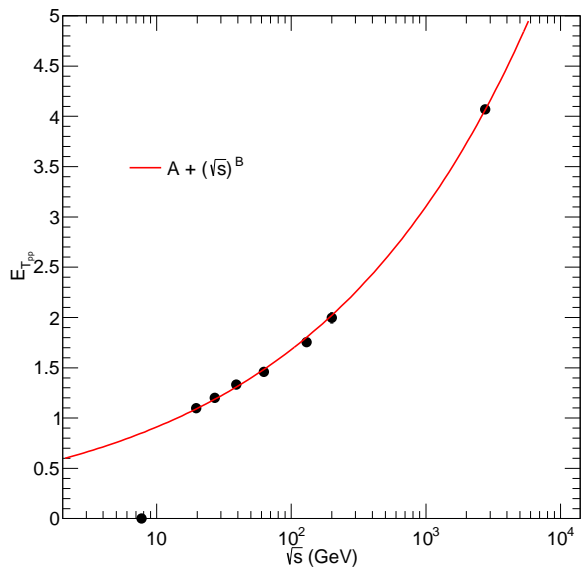}
\caption{The variation of $E_{T_{pp}}$ as a function of center of 
mass energy. This is parameterized by a power law shown by the solid 
line.}
\label{f9}
\end{figure}

\begin{figure}
\includegraphics[scale=0.4]{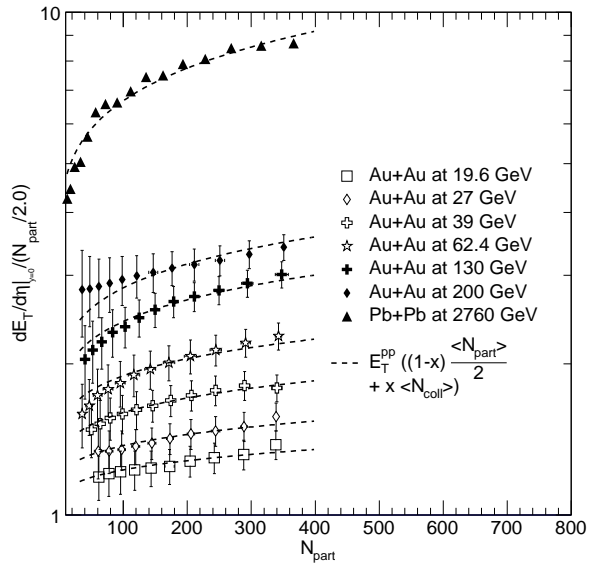}
\caption{Charged particle transverse energy densities scaled by 
the number of participants measured in heavy ion collisions at  
RHIC (Au+Au)~\cite{phenixpaper} and LHC (Pb+Pb)~\cite{alice3} energies. The dashed line represents the fits to Eq.~\ref{e1}.}
\label{f10}
\end{figure}

\begin{figure}
\includegraphics[scale=0.4]{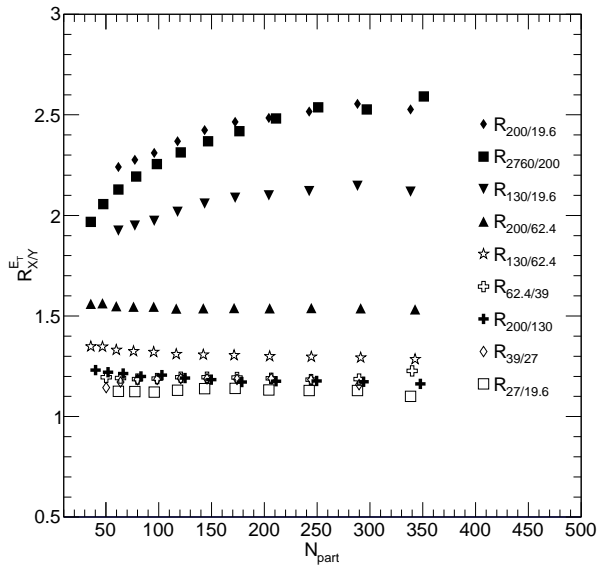}
\caption{
The variation of the double ratio of the 
charged particle energy density (refer text) as a function of $\langle N_{part} \rangle$.}
\label{f11}
\end{figure}

The measured transverse energy density, $dE_{T}/d\eta$ as a function of 
$\langle N_{part} \rangle$ is studied by the two-component 
model~\cite{twocomp} approach described in previous sections. As discussed in the
method for $dE_{T}/d\eta$, the $x$ parameter  values were used from
the fits of centrality dependence of  $dN_{ch}/d\eta$ while
$E_{T_{pp}}$  was treated as a free parameter. 
The results of the fit are shown in Figure~\ref{f8}.  The model 
describes the data very nicely for all energies from RHIC to LHC. 
The extracted values of  $E_{T_{pp}}$ is plotted as a function of $\sqrt{s}$ 
in Figure~\ref{f9}. The values are  parameterized with the same function
as that of  $n_{pp}$  as depicted in Figure~\ref{f9}.
Figure~\ref{f10} depicts the scaled transverse energy densities as a
function of centrality for various energies.  The scaled energy
densities do not have any centrality dependence.
This behaviour is also seen in the pair wise ratios of $dE_{T}/d\eta$ 
at two different set of energies as shown in Figure~\ref{f11}. 
A variable $R^{E_{T}}_{X/Y} \, = \frac{dE_{T}/d\eta}{\langle N_{part}\rangle/2}|_X/\frac{dE_{T}/d\eta}{\langle N_{part}\rangle/2}|_Y
$,  is used to denote the magnitude of the  ratio of transverse energy
densities at energy $X$ and energy $Y$. 
The ratio shows a flat behavior as a function of system size for 
all the collision energy ranges considered.  The magnitude of the
ratios depends on the  factor of difference in energy. For example,  the values of
$R_{2760/200}$ is comparable to that of $R_{200/19.6}$ as the
collision energy  difference is  more or less of same order. The same
was not observed for case of the multiplicity densities. The values of
$R_{2760/200}$ is consistently higher than $R_{130/19.6}$ inspite of the
$x$ factor being higher in the later. This was not seen in the trend  of
pairwise ratios of multiplicity densities where both the ratios were
comparable.
 One can conclude that the role of $x$ is not significant in dictating
 the observed trend in case of transverse energy densities.

\section{Summary}
The midrapidity charged-particle multiplicity density and the
transverse energy density  measurements were  described nicely by the
two-component Glauber approach for a broad range of energies.
The analysis shows that the fractional contribution of ``hard processes"
parametrized by the value of $x$ increases with an increase in beam
energy.  A  centrality dependence of the same is also shown but the 
increase is not very significant with respect to centrality.
The pair wise ratios of the multiplicity densities at two different
energies shows a scaling which  depends on the  interplay of energy 
 difference and the contribution from the hard scattering component.
The  later plays no role in the observed order of ratios for the
transverse energy densities. Although the model is bit crude and
simplistic  in its approach, the  description of the the overall
features of the  centrality dependence of multiplicity and
transverse energy densities at various energies is in good agreement
with the data.


\noindent
\begin{thebibliography}{50}
\medskip
\bibitem{twocomp} D. Kharzeev and M. Nardi, Phys. Lett. B{\bf507}, 121
(2001).

  
\bibitem{mitpaper} L. Zhou and G. S.F. Stephans, Phys. Rev. C{\bf 90}, 014902 (2014).

\bibitem{sps1} M. C. Abreu {\it et al.}, (NA50 Collaboration), Phys. Rev. Lett. B{\bf 530}, 43 (2002). 

\bibitem{sps2}  M. M. Aggarwal {\it et al.}, (WA98 Collaboration), Eur. Phys. J. C{\bf 18}, 651 (2001). 

\bibitem{sps3}  F. Antinori {\it et al.}, (NA57 and WA97 Collaboration), J. Phys. G{\bf 27}, 391 (2001). 

\bibitem{alice1} K. Aamodt {\it et al.}, (ALICE Collaboration), Phys. Rev. Lett. {\bf 106}, 032301 (2011).

\bibitem{phenixpaper} A. Adare {\it et al.}, (PHENIX Collaboration), Phys. Rev. C{\bf 93}, 024901 (2016). 

\bibitem{alice2} J. Adam {\it et al.}, (ALICE Collaboration), Phys. Rev. Lett. {\bf 116}, no.22, 222302 (2016).
\bibitem{alice3} C. Loizides for ALICE Collaboration, J. Phys. 
G{\bf 38}, 124040 (2011).
\bibitem{glauber} M. M. Miller {\it et al.}, Ann. Rev. Nucl. Part. Sci. {\bf 57}, 205-243 (2007).
\bibitem{npp_rhic} B. B. Back {\it et al.}, (PHOBOS Collaboration), Phys. Rev. C{\bf 70}, 021902 (2004).
\bibitem{npp_lhc1} J. Adam {\it et al.}, (ALICE Collaboration), arXiv:1509.07541[nucl-ex].
\bibitem{npp_lhc2} J. Adam {\it et al.}, (ALICE Collaboration), Phys. Lett. B{\bf 753}, 319-329 (2016).




\end{thebibliography}
\end{document}